\documentclass[prl,twocolumn,aps,amssymb,footinbib]{revtex4}
\usepackage{amssymb}
\usepackage{graphicx}
\usepackage{amsmath}
\usepackage{times}
\usepackage{color}
\usepackage{subfigure}
\usepackage{setspace}
\usepackage{bm}

\newcommand {\beq} {\begin{equation}}
\newcommand {\eeq} {\end{equation}}
\newcommand {\bqa} {\begin{eqnarray}}
\newcommand {\eqa} {\end{eqnarray}}

\newcommand {\ep} {\ensuremath{\epsilon}}

\newcommand {\cd} {\ensuremath{c^\dagger}}
\newcommand {\ca} {\ensuremath{c^{\phantom{\dagger}}}}

\newcommand {\gd} {\ensuremath{\gamma^\dagger}}
\newcommand {\ga} {\ensuremath{\gamma^{\phantom{\dagger}}}}

\newcommand{\kk} {\ensuremath{{\bf k}}}

\newcommand {\qq} {\ensuremath{{\bf q}}}

\begin{document}

\begin{abstract}
  We propose a pump probe experiment for detecting the d-wave
  superfluid and d-density wave phases of ultracold Fermions on an
  optical lattice. The pump consists of periodic modulations of the
  optical lattice intensity which creates quasiparticle pairs in these
  systems. The changes in the momentum distribution under the drive
  can be used to measure quasiparticle dispersion and gap anisotropy.
  Further, we show that the pattern of peaks and dips in the spin
  selective density-density correlation function provides a phase
  sensitive probe of the symmetry of the order parameter in these
  systems.
 
\end{abstract}

\title{Detecting d-wave superfluid and d-density wave states of ultracold Fermions on optical lattices}

\author{David Pekker, Rajdeep Sensarma and Eugene Demler}  
 \affiliation{ Physics Department,
Harvard University, Cambridge, Massachusetts 02138, USA}
\maketitle

Ultracold atoms on optical lattices are emerging as candidates for
building a {\it quantum simulator}~\cite{Feynman1982} of various
lattice models used to study strongly interacting many body systems.
These systems have already been used to implement important lattice
Hamiltonians like the Bose Hubbard model~\cite{Greiner2002}, which is
central to the theory of superfluid-insulator
transitions~\cite{Fisher1989} and attractive and repulsive Fermi
Hubbard models, which are central to the study of strongly interacting
superfluids~\cite{Randeria} and high temperature
superconductors~\cite{Anderson1987} respectively. The recent
observation of Mott insulating states in the repulsive Fermi Hubbard
model~\cite{Jordens2008, Schneider2008} shows a novel manifestation of
strong interaction in these systems.

The attractive feature of cold atom systems is that they provide a
clean setting with controlled and tunable access to the strongly
interacting parameter regime of model Hamiltonians. However, a good
quantum simulator also requires methods to extract information about
the many body correlations in the system.  This is specially relevant
for simulating strongly correlated systems with complex phase
diagrams, high $T_c$ superconductors, where the quantum simulation can
shed light on the nature of the various phases and test the various
competing theories that have been proposed. A quantum simulator which
aims to choose between competing theoretical descriptions should be
able to probe the characteristic correlations and excitations of the
various observed and proposed phases.

In the setting of strongly correlated system condensed matter setting
like high $T_c$ superconductors, measurement of thermodynamic
quantities like specific heat and of transport properties like
resistivity, Hall conductivity, and Nerst effect as well as various
spectroscopic techniques like neutron scattering and optical
conductivity have been used to identify the various thermodynamic
phases and have yielded a lot of information about correlations and
excitations. Detailed information about the electronic structure
obtained from ARPES~\cite{ARPESreview} and STM~\cite{Hoffman2002,
  McElroy2003} measurements has been crucial in identifying the nature
of these thermodynamic phases. A crucial aspect of constructing the
phase diagram is to identify the non-trivial symmetry of the various
order parameters e.g. the d-wave symmetry of the superconductor.  The
techniques mentioned above contain indirect information about the
symmetry, but a more reliable method for probing the symmetry of the
superconducting order parameter has been to perform a phase-sensitive
measurement, i.e. look for a $\pi$ phase-shift in the corner junction
SQUID~\cite{Wollman1993}. However, this method is not directly
applicable to the cold atom setting. Hence, it is important to develop
new experimental techniques for cold atom systems to probe
the correlations and order parameter symmetry.

In this Letter, we consider the problem of characterizing d-wave
Fermionic superfluids and states with d-density wave order. The
d-density wave state has been proposed as a possible candidate for the
pseudogap phase~\cite{Chakravarty2001}. We focus on two issues:
(1)~determining the quasi-particle dispersion and identifying the
location of the nodes in the gap function if there are any and
(2)~experimentally identifying the order parameter symmetry. Our
method consists of looking at the response of the system to periodic
modulation of the optical lattice intensity.  This drive creates pairs
of quasi-particles of opposite momenta predominantly with total energy
corresponding to the frequency of the drive. The change in momentum
distribution $n_k$ in response to the drive can be used to measure the
quasiparticle dispersion.  Further, the pattern of peaks and dips in
different density-density correlation functions, which result from the
interference of the quasiparticles excited by the drive, encode the
symmetry of the order parameter (e.g. s-wave vs. d-wave pairing).  The
momentum distribution can be measured by a ballistic expansion
experiment, while the correlators can be measured by noise
correlations in the real space density profiles.  The Letter consists
of two parts: in the first part, we describe our method and apply it
to the d-wave superfluid case, and in the second part we apply it to
the d-density wave case.

\begin{figure}
\includegraphics[width=6cm]{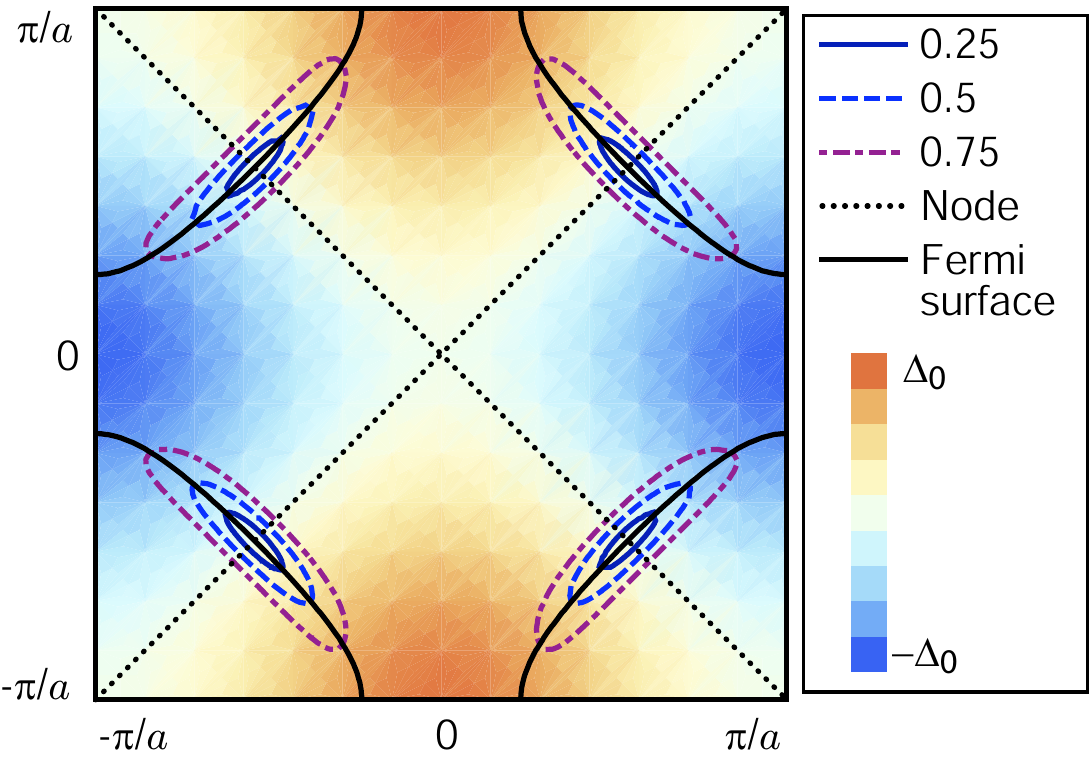}
\caption{Brillouin zone of a tight binding d-wave BCS
superfluid. Solid black lines indicate the location of the Fermi
surface (without the pairing interaction), while the dotted black
lines indicate the location of the nodes. The shading indicates the
value of order parameter $\Delta_\kk$. Several equal quasiparticle energy
contours, {\it bananas}, are shown with colored lines. }
\label{Fig:banana}
\end{figure}

{\it d-wave Superfluid\/} --- We will use the simplified BCS
description of a d-wave superfluid, which helps in demonstration of
our proposed methods; however the qualitative results that we obtain
are independent of the microscopic description, and the non-BCS nature
of the superfluid will only result in quantitative discrepancies. Thus
this method may be applied to the superfluid phase of the
Fermi-Hubbard model (assuming that it exists). We start with the tight
binding BCS Hamiltonian on a two dimensional square lattice, suitably
generalized for arbitrary pairing symmetry,
\begin{align}
H_0=\sum_{\kk} 
\left(\begin{array}{cc} 
\cd_{\kk\uparrow} & \ca_{-\kk\downarrow} 
\end{array} \right)
\left(\begin{array}{cc} 
\xi_\kk & \Delta_\kk \\
\Delta_\kk & -\xi_\kk
\end{array} \right)
\left( \begin{array}{c} 
\ca_{\kk\uparrow} \\
\cd_{-\kk\downarrow} 
\end{array} \right)
\end{align}
where $\cd_{\kk\sigma}$ is the creation operator for an atom with
flavor $\sigma \in \{\uparrow,\downarrow\}$ and momentum $\kk$ and
$\xi_\kk=\ep_\kk-\mu$ is the tight binding dispersion with $\mu$ the
chemical potential and $\ep_\kk=-2J(\cos k_x +\cos k_y)$ where $J$ is
the tunneling matrix element. The mean field $\Delta_\kk$ describes
the pairing interaction, and its $\kk$ dependence reflect the pairing
symmetry. We focus on the case of d-wave pairing
$\Delta_\kk=\Delta_0(\cos k_x-\cos k_y)$.

We drive the system by periodic modulation of the optical lattice
intensity $V(t)=V_0+\delta V \sin(\omega t)$, which modulates the
tunneling matrix element $J\rightarrow J(1+g_1(t))$.  The modulation
of the superconducting gap $\Delta_0 \rightarrow \Delta_0(1+g_2(t))$
depends on the microscopic mechanism of
pairing. However, the precise form of $g_2(t)$ is not
important for qualitative results; the important point is that both
$g_1(t)$ and $g_2(t)$ have similar time dependence ($\sim \sin
(\omega t)$)~\cite{check}. (For the purposes of the example calculations we set
$g_2(t)=2g_1(t)=2 \lambda \sin(\omega t)$, where
$\lambda\sim\sqrt{V_0/E_r} \delta V_0/V$. This choice is
motivated by the expectation that $\Delta\sim
J^2/U$~\cite{Anderson1987}.)

For illustrative purposes, we shall consider driving a system that
starts at zero temperature, i.e.~with the initial condition that all
quasiparticle states are empty. As the perturbation Hamiltonian only
creates or destroys quasiparticles in pairs, we truncate the Hilbert
space to exclude states with unpaired quasiparticles. This truncated
Hilbert space is spanned by wavefunctions of the BCS form
\begin{align}
\Psi(t) = \prod_\kk
\left(u_\kk(t)+v_\kk(t) \cd_{\kk\uparrow}\cd_{-\kk
\downarrow} \right)|0\rangle. \label{eq:PsiBCS}
\end{align}
The initial state corresponds to the BCS solution
$u_{\kk}=\text{sign}(\Delta_\kk)\left[
\frac{1}{2}\left(1+\xi_\kk/E_\kk\right)\right]^{1/2}$ and 
$v_{\kk}=
\left[\frac{1}{2}\left(1-\xi_\kk/E_\kk\right)\right]^{1/2}$ with 
$E_\kk=\sqrt{\xi_\kk^2+\Delta_\kk^2}$ the quasiparticle dispersion 
of the unperturbed state. The equations of motion for the coherence 
factors are
\begin{align}
i \partial_t\left(
\begin{array}{c}
u_{\kk}\\
v_{\kk}
\end{array}
\right)
=
\left(
\begin{array}{c c}
0 & \Delta_\kk(t) \\
\Delta_\kk(t) &  2 (\xi_\kk + g_1(t) \epsilon_\kk)
\end{array}
\right)
\left(
\begin{array}{c}
u_{\kk}\\
v_{\kk}
\end{array}
\right).
\end{align}
where $\Delta_\kk(t)=\Delta_\kk(1+g_2(t))$.
In describing the evolution, we do not maintain the self-consistency equations 
for $\Delta$ and $\mu$. This is justified in the linear response regime where 
we do not create a lot of quasiparticles.

To gain a better understanding of the dynamics, we use second order
perturbation theory in $\lambda$ to study the evolution of the system.
It is more convenient to perform perturbation theory calculations in
the quasiparticle basis
\begin{align}
\Psi(t) = \prod_\kk \left(\psi_{1,\kk}(t) + \psi_{2,\kk}(t)
\gd_{\kk\uparrow}\gd_{-\kk \downarrow} \right) |\Psi_\text{BCS}\rangle,
\end{align}
where $\gd_{\kk,\sigma}=u_\kk \cd_{\kk,\sigma} +\sigma v_\kk
\ca_{-\kk,\bar{\sigma}}$ are the quasiparticle creation operators.
In the tight binding d-wave superfluid, the contours of 
equal quasiparticle energy look like {\it bananas} (See Fig ~\ref{Fig:banana}).
When $\kk$ is off resonance (i.e.~away from the banana contour
corresponding to $\omega=2E_\kk$), with detuning $\tilde{\omega}_\kk=2
E_\kk - \omega$, $|\psi_{2,\kk}|$ oscillate between $0$ and $\sim 
\lambda u_\kk v_\kk \epsilon_\kk / \tilde{\omega}_\kk$ at
approximately the drive frequency. On the other hand, if $\kk$ is on
resonance, then $|\psi_{2,\kk}|$ oscillate between $0$ and $1$ with
the period $P_\kk \sim (2 \lambda u_\kk v_\kk
\epsilon_\kk)^{-1}$. Thus, after the system has been driven for some
time, the quasi-particle momentum distribution function $\langle
\gd_\kk \ga_\kk \rangle$ will develop peaks along the banana contours
corresponding to $\omega=2E_\kk$, with maxima near the banana tips. To
optimize the peak height, the system should be driven for a time $t
\sim P_\kk^{-1}$ for $\kk$ near a banana tip.
For the remainder of the paper, we shall solve the equations of motion
numerically, but use the insights from this simple perturbation theory
to qualitatively explain the results.

\begin{figure}
\includegraphics[width=8cm]{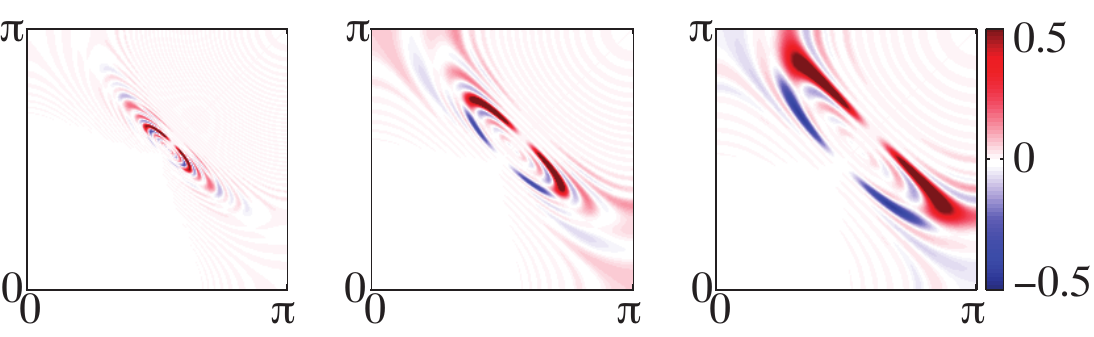}
\caption{Change in the momentum distribution function under the
  influence of the drive $\langle n_k(t=25/J) \rangle-\langle n_k(0)
  \rangle$ for a BCS d-wave superfluid, with $\Delta_0=0.5 J$, for
  various drive frequencies: $\omega=\Delta_0/2$ (left), $\Delta_0$
  (center), $1.5 \Delta_0$ (right). }
\label{Fig:nk}
\end{figure}

{\it Momentum distribution function $n_k$\/} --- To extract
information about the dispersion relation, we propose to look at the
momentum distribution of the atoms, $n_k = \langle \cd_\kk \ca_\kk
\rangle = 2 |v_\kk(t)|^2$. Experimentally, $n_k$ may be obtained from
a band-mapping free expansion experiment~\cite{FreeExp}.
As an illustration, we consider the case of a 
d-wave superfluid ($\mu=4.5$, $\Delta_0=1.0$, $J=1$) that we drive at
three different frequencies $\omega=\{0.5, 1.0, 1.5\} \Delta_0$. The 
change in momentum distribution at a time $t=25$ is plotted in Fig~\ref{Fig:nk}.
As explained before, at a given frequency $\omega$, the excitations
are being driven predominantly on the banana-shaped contour
corresponding to $\omega=2E_\kk$.  This leads to large changes in the
momentum distribution along the same contour.  The dispersion can then
be extracted by varying the drive frequency. If the frequency
decreases (increases), the banana-shaped contour shrinks (grows) as
can be seen in the left (right) panel of Fig.~\ref{Fig:nk}. The
fringes correspond to the diffraction pattern that is formed at short
times. The dispersion will reveal the $\kk$ dependent gap function and
hence the d-wave nature of the superfluid. In this respect, this
method is similar to the ARPES determination of the momentum
dependence of the gap in high $T_c$
superconductors~\cite{Campuzano1999}.

In this context we will briefly comment on the response of a s-wave
superfluid to the same drive. Since the s-wave system is gapped, one
would not observe any response before the drive frequency exceeds
twice the gap. In contrast, the low energy nodal quasiparticles in a
d-wave superfluid lead to response at all drive frequencies.

\begin{figure}
\includegraphics[width=8cm]{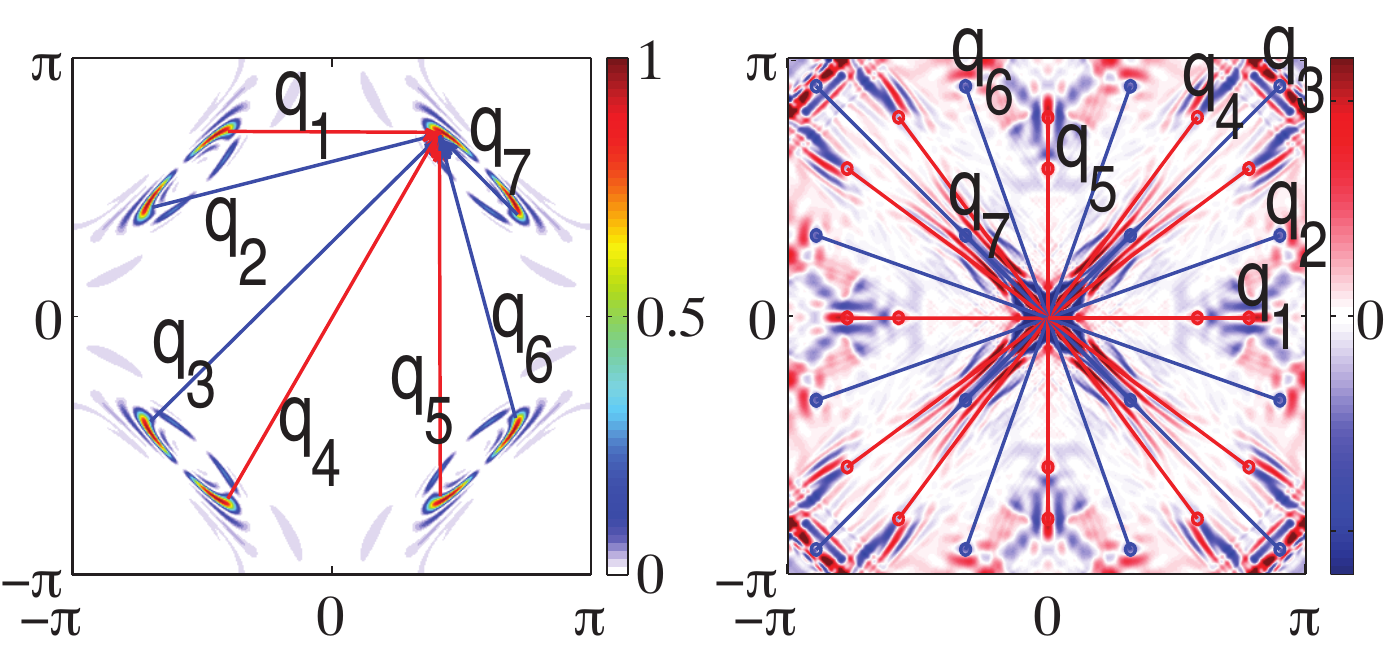}
\caption{Left: Quasiparticle density $|\psi_{\kk,2}(t=25/J)|^2$. Red (blue) arrows
indicate wave-vectors that connect tips of bananas between which  $\Delta_\kk$ 
does not change (changes) sign.
 Right:
Correlation function $\langle \rho_\uparrow(q) \rho_\downarrow(-q)
\rangle_{t}-\langle \rho_\uparrow(q) \rho_\downarrow(-q) \rangle_{0}$
for a d-wave superfluid.
Pairing symmetry is
encoded in peaks (red) and dips (blue) that develop in the correlation
function as described in the text. (for both panels
$\omega=\Delta_0$)}
\label{Fig:rhorho}
\end{figure}

{\it Spin up--spin down density correlation function } --- Next, we
show that the changes in the spin up--spin down density correlation
function $N^{\uparrow \downarrow}(q)=\langle
\rho_\uparrow(q)\rho_\downarrow(-q) \rangle$ show a pattern of peaks
and dips which reveals the order parameter symmetry in a way similar
to corner junction SQUID measurements in condensed matter systems.
This correlation function can either be measured using elastic light
scattering or extracted from the noise in real space measurement of
the particle density. The point of measuring this particular
correlation function is that it is sensitive to the symmetry of the
pairing interactions in the system.

We can relate $N^{\uparrow \downarrow}(q)$ to the BCS wavefunction at
time $t$ defined by the coherence $u_\kk(t)$ and $v_\kk(t)$ as
follows
\begin{align}
N^{\uparrow \downarrow}(q)&= \sum_{\kk,\kk'} 
\left\langle
c_{\kk+\qq \uparrow}^\dagger(t) c_{\kk\uparrow} (t) c_{ \kk'-\qq \downarrow}^\dagger(t) c_{\kk' \downarrow} (t)
\right\rangle \\
&=  \sum_{\kk} 
\left\langle 
c_{\kk+\qq \uparrow}^\dagger(t) c_{-\kk-\qq \downarrow}^\dagger(t)
\right\rangle
\left\langle 
c_{-\kk \downarrow} (t)  c_{\kk\uparrow} (t)
\right\rangle\\
&=
\sum_\kk u_\kk(t) v^*_\kk(t)
u^*_{\kk+\qq}(t) v_{\kk+\qq}(t).
\end{align}
Here, in the middle step, we have used the fact that for the BCS
wavefunction only the anomalous pair correlations are non-zero.  We
note that the d-wave symmetry of $u_\kk v^\ast_\kk$ is preserved
throughout the time evolution of the system. The form of $N^{\uparrow
  \downarrow}(q)$ looks like the product of Cooper pair wavefunctions
at $\kk$ and $\kk+\qq$, and therefore measures the interference of
quasiparticles created at the respective wavevectors.  Driving creates
predominantly creates quasiparticles near the tips of the resonant
banana contours.  The interference of these quasiparticles result in
peaks or dips in the noise correlation function whenever the
wave-vector $\qq$ that appears in $N^{\uparrow \downarrow}(q)$
connects a pair of banana tips located at $\kk$ and $\kk+\qq$.

The change of $N^{\uparrow \downarrow}(q)$ after driving the system
for $\sim 4$ cycles is depicted in the right panel of
Fig.~\ref{Fig:rhorho} for the case of the d-wave superfluid. To
decipher the meaning of the various peaks and dips we compare the
right panel of Fig.~\ref{Fig:rhorho} to the left panel which depicts
the corresponding change in the quasiparticle distribution function
$\langle \gamma^\dagger_\kk \ga_\kk \rangle$. Connecting the tips of
the bananas (or the peaks in the quasiparticle density) with arrows in
all possible ways we identify 7 unique wave-vectors, which are labeled
$q_1$ through $q_7$ in Fig.~\ref{Fig:rhorho}. The vectors that connect
the remaining pairs of tips may be obtained by reflections of
$q_1..q_7$ across the nodal directions.  The presence of a peak/dip at
a given $\qq$ vector depends on the time of drive; however the
relative pattern of peaks and dips is governed by the symmetry of the
order parameter. If the wave vector connecting two banana tips where
$\Delta_\kk$ and $\Delta_{\kk+\qq}$ has the same sign (say $\qq_1$)
has a peak in the signal, the wave-vectors connecting the tips where
the order parameter has opposite signs (say $\qq_2$) should have a dip
and vice versa.  Note that for s-wave pairing where order parameter
does not change sign we would have all peaks or all dips depending on
time of drive. For other singlet pairing symmetries the relative
pattern would change depending on the location in the Brillouin zone
where the order parameter changes sign.

Therefore, given the noise correlation function we can (1) determine
the dispersion relation along the Fermi surface by measuring how the
distances between the peaks and the dips varies with the drive
frequency; (2) determine the pairing symmetry by figuring out the
relative sign of $\Delta_\kk$ at the various banana tips from the
pattern of dips and peaks. 

The important point to note is that our method asks a binary question
of whether there is a peak or a dip much like corner junction SQUID
experiment. In this sense this method is robust to quantitative
changes in the signal due to strong correlation effects.

{\it d-Density Wave\/} --- Next, we show that similar methods can be
applied to the d-density wave states which have been proposed as
prospective candidates for the pseudogap phase of high $T_c$
superconductors~\cite{Chakravarty2001}.  In this state the operator
$\cd_\kk \ca_{\kk+\qq}$ acquires an expectation value $\Delta_\kk$. We
concentrate on the commensurate case in which the ordering occurs at
momentum $\qq=(\pi,\pi)$. At the mean field level the system is
described by the Hamiltonian
\begin{align}
H_0=\sum_{\kk}\xi_\kk \cd_{\kk}\ca_{\kk} +
\Delta_\kk \cd_{\kk}\ca_{\kk+\qq} +
\Delta_\kk^*\ca_{\kk}\cd_{\kk+\qq}, \label{Eq:DDW}
\end{align}
where the mean field has the same form as for the d-wave superfluid
case $\Delta_\kk=\Delta_0 (\cos(k_x)-\cos(k_y))$. The commensurate
ordering halves the Brillouin zone but fills it with two different (quasiparticle) 
bands. The two bands touch at the points $(\pm \pi/2, \pm \pi/2)$. At
half-filling ($\mu=0$), the lower band is completely filled and the
upper band is completely empty. Away from half filing there are empty 
states in the lower band. Optical lattice modulation excites
quasiparticles from the lower branch to the upper branch leaving
quasiholes behind in the lower band. In other words, in analogy with
the d-wave superfluid, the drive creates quasihole-quasiparticle pairs
with zero total momentum and total energy $\omega$. The wave function in analogy to
Eq.~(\ref{eq:PsiBCS}), can be written in the form
\begin{align}
\Psi(t) = \prod_\kk \left(\alpha_\kk(t) \cd_\kk + \beta_\kk(t)
\cd_{\kk+\qq} \right). \label{eq:PsiDDW}
\end{align}

The dispersion relation for quasiparticles can again be obtained by
looking at the response of the momentum distribution to the
drive. Because of the folding, the momentum distribution for the inner
half-zone and outer half zone must be patched together
\begin{align}
n_\kk=\left\{\begin{array}{cc}
|\alpha_\kk(t)|^2 & \quad\quad\quad \kk \in \text{inner zone}\\
|\beta_\kk(t)|^2 & \quad\quad\quad \kk \in \text{outer zone}
\end{array} \right.
\end{align}
There is one important difference in that tuning the system
away from half-filling does not shift the location of the bananas as
the density wavevector $\qq$ has been assumed to be fixed. So
independent of filling, the centers of the bananas remain fixed at
$(\pm \pi/2, \pm \pi/2)$. Away from half-filling the low frequency
response (inner portion of the banana) disappears as both branches are
either empty or filled and thus a quasihole-quasiparticle pair cannot
be created.

\begin{figure}
\includegraphics[width=8cm]{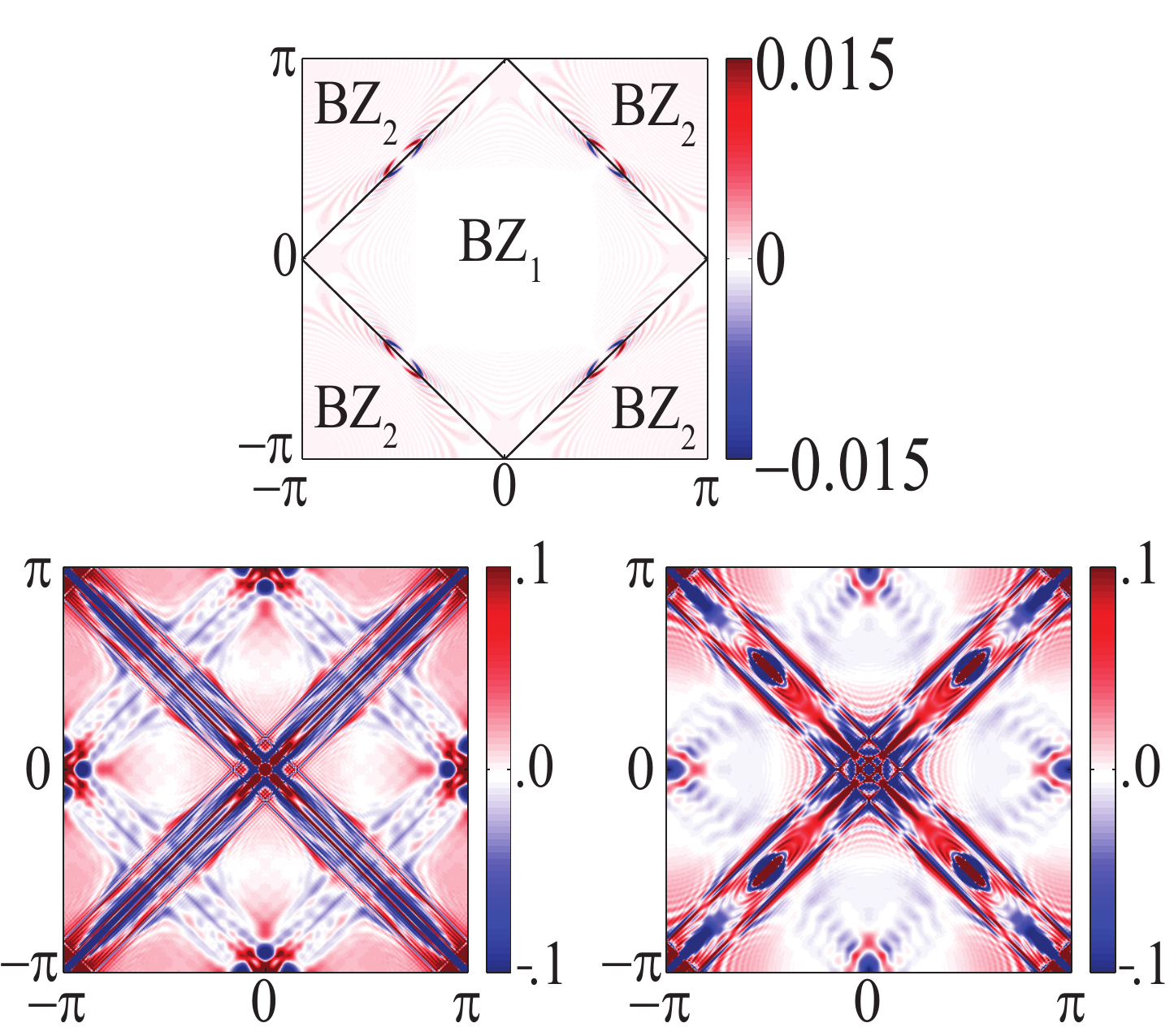}
\caption{Effects of drive at frequency $\omega=\Delta_0/2$ on density
  wave states. Change in the momentum distribution function $\langle
  n_\kk(t=25/J) \rangle-\langle n_\kk(0) \rangle$ for d-density-wave
  at half-filling (top). Change in the density-density correlation
  function $\langle \rho_\kk\rho_{\qq-\kk} (t=25/J) \rangle-\langle
  \rho_\kk\rho_{\qq-\kk} (t=0) \rangle$ for anisotropic s-density-wave
  $\Delta_\kk=0.5 |\cos(k_x)-\cos(k_y)|$ (left) and d-density-wave
  $\Delta_\kk=0.5 \cos(k_x)-\cos(k_y)$ (right).  }
\label{Fig:rhorhoDDW}
\end{figure}

The symmetry of the order parameter is encoded in the pattern of peaks
and dips of the $\langle \rho_\kk \rho_{\qq-\kk} \rangle$. We compare
this pattern for two cases $\Delta_\kk=\Delta_0 (\cos(k_x)-\cos(k_y))$
and $\Delta_\kk=\Delta_0 |\cos(k_x)-\cos(k_y)|$. Because the peaks in
the quasiparticle density near the tips of the bananas are not as
sharp as for the case of d-wave superfluid it is hard to identify the
origin of various peaks and dips, but there is a clear distinction
between the two cases.

{\it Concluding remarks\/} --- Probing the properties of the
correlated state of ultracold atoms is an important part of building a
``quantum simulator.'' In this Letter we propose to use periodic
modulation of the optical lattice amplitude to probe Fermionic
superfluids and density wave states. This probe is relevant to
simulating the repulsive Hubbard model which is thought to be a
minimal model of high $T_c$ superconductors. The methods we suggest
are useful for probing states with nodes or at least an anisotropic
gap.  We have shown that: First, we can determine the quasi-particle
dispersion and gap anisotropy by looking at the change in the momentum
distribution created by the drive. Second, by measuring the
appropriate density-density correlation function we can determine the
pairing symmetry from the pattern of peaks and dips that develops. By
measuring the distances between the peaks and dips, we can also
extract the gap anisotropy along the Fermi-surface from the second
method method. 

{\it Acknowledgements\/} --- It is our pleasure to thank R. Barankov
for discussions that helped to identify the problem and M. Lukin,
M. Greiner, and W. Zwerger for useful discussions.

These authors acknowledge the support of DARPA, CUA, MURI, and NSF
grant DMR 0705472. R.S. and D.P. were partially supported by NSF grant
DMR-05-41988.


\begin{thebibliography}{99}
\bibitem{check}
We have explicitly checed that the altering the ratio $g_1/g_2$ does not qualitatively effect the patterns in Figs.~\ref{Fig:nk} and \ref{Fig:rhorho}.
\end{thebibliography}
\end{document}